\documentclass[aps, prl, twocolumn, nofootinbib, showpacs]{revtex4-1}
\usepackage{amsfonts,amssymb,amsmath}
\usepackage{mathrsfs}
\usepackage{amsthm}
\usepackage{graphics,graphicx,epsfig}
\usepackage[dvipsnames]{xcolor}
\usepackage[normalem]{ulem}
\usepackage{algorithm}
\usepackage{algorithmicx}
\usepackage{algpseudocode}
\usepackage{multirow}

\usepackage{color}

\usepackage{epstopdf}

\begin{document}

\title{Tomographic Witnessing and Holographic Quantifying of Coherence
}
\author{Bang-Hai Wang$^{1}$}
\email{bhwang@gdut.edu.cn}
\author{Si-Qi Zhou$^{2,3}$}
\email{sqzhou@sjtu.edu.cn}
\author{Zhihao Ma$^{2,3}$}
\email{mazhihao@sjtu.edu.cn}
\author{Shao-Ming Fei$^{4,5}$}%
\email{feishm@cnu.edu.cn}
\affiliation{%
$^1$School of Computer Science and Technology, Guangdong University of Technology, Guangzhou 510006, China\\
$^2$School of Mathematical Sciences, MOE-LSC, Shanghai Jiao Tong University, Shanghai 200240, China\\
$^3$Shenzhen Institute for Quantum Science and Engineering, Southern University of Science and Technology, Shenzhen 518055, China\\
$^4$School of Mathematical Sciences, Capital Normal University, Beijing 100048, China\\
$^5$Max-Planck-Institute for Mathematics in the Sciences, 04103 Leipzig, Germany
}%


\begin{abstract}
The detection and quantification of quantum coherence play significant roles in quantum information processing. We present an efficient way of tomographic witnessing for both theoretical and experimental detection of coherence. We prove that a coherence witness is optimal if and only if all of its diagonal elements are zero. Naturally, we obtain a bona fide homographic measure of coherence given by the sum of the absolute values of the real and the imaginary parts of the non-diagonal entries of a density matrix, together with its interesting relations with other coherence measures like $l_1$ norm coherence and robust of coherence.

%


\end{abstract}


\maketitle

\textit{Introduction -}
Linear superposition lies at the heart of quantum mechanics. It is the most fundamental and significant principle of quantum physics and lays down the quantum parallel computing and quantum communication theory.
As a characterization of liner superposition, quantum coherence has attracted considerable interest and has rapidly grown into a hot research field. The quantification, characterization, manipulation, dynamical evolution, and operational application of coherence have been widely investigated (see, e.g.,  \cite{Streltsov17,Hu2018,Huang2017} and references therein).

Many methods and tools used in investigating quantum entanglement have been employed in
dealing with coherence. Inspired by the entanglement witnesses \cite{Augusiak11}, the notion of coherence witnesses was first introduced by Napoli et al. \cite{Napoli16,Piani16}. They found that the expectation value of a coherence witness operator for any quantum state provides a quantitative lower bound to the robustness of coherence, which can be endowed with an operational interpretation in terms of channel discrimination \cite{Napoli16,Piani16}. A coherence witness $W$ is an Hermitian operator such that (i) $\text{tr}(W\delta)\geq0$ for all incoherent states $\delta$, (ii) there exists at least one coherent state $\pi$ such that $\text{tr}(W\pi)<0$,  and (iii) $tr(W)=1$ for normalization. The condition (i) could also be defined by (i') $\text{tr}(W\delta)=0$ for all incoherent states $\delta$. However, condition (i) is claimed to be experimentally friendlier than condition (i') \cite{Napoli16,Piani16}. Tightening condition (ii) to (ii'), $\text{tr}(W\pi)\neq0$, Ren et al. put forward the so-called stringent coherence witness which satisfies conditions (i') and (ii'). The optimal witness $W$ is defined to be the one such that $\text{tr}(W\pi)$ attains the maximal value \cite{Ren2017}. Recently, some experiments and applications involved in coherence witnesses have also been presented \cite{Wang17,Nie19,Zheng18,Ringbauer18,Ma19} based on coherence witnesses satisfying conditions (i) and (ii).

Quantifying coherence is essential for both quantum foundations and quantum technologies \cite{Piani16,Napoli16}. A plethora of measures have been presented. The $l_1$ norm  measure of coherence and the relative entropy measure of coherence \cite{Baumgratz14}, the distillable coherence and coherence of formation \cite{Yuan2015,Fang2018,Winter2016}, entanglement \cite{Streltsov2015}, the coherence concurrence \cite{Du2015,Qi2017}, the robustness of coherence \cite{Napoli16}, the max-relative entropy of coherence \cite{Bu2017}, Hellinger distance \cite{Jin2018}, Logarithmic coherence number \cite{Xi2019}  etc. have been employed to quantify coherence.
However, most of the measures are for given known states or require state tomography by experiments.


In this paper, we develop a theoretical groundwork for coherence witnesses and apply it to the experimental verification and quantification of quantum coherence in a quantum optical setting. Inspired by the optimization of entanglement witnesses \cite{Lewenstein00}, we characterize a coherence witness by using a set of coherent states detected by the coherence witness. We introduce a new definition of optimal coherence witness and show that a coherence witness $W$ is optimal if and only if its diagonal elements are zero. We find a group of optimal coherence witnesses which just corresponds, one by one, to the generators of the standard special unitary group for non-diagonal elements. We employ state tomography method to realize the optimal (traceless) coherence witnesses.
The state-tomography method generally contains redundant measurements for coherence detection, as one does not need to know the full information about a quantum state to judge if the state is coherent. Our method employs both the theory of optimal coherence witnesses and the tomographic technology in experiment. It provides a more economic way to detect coherence than the usual coherence witness in experiments, in analogy to the theory of entanglement witnesses \cite{Guhne09,Horodecki09}. Naturally, we put forward a bona fide coherence measure of coherence, which takes over all real and imaginary parts of all non-diagonal elements of a density matrix. We further characterize this measure and reveal its relations with the $l_1$ norm measure and the robustness of coherence.


\textit{Optimal coherence witnesses -} The coherence of a state is defined with respect to a fixed basis $\{|i\rangle\}_{i=1,2,\cdots,d}$ of the related $d$-dimensional Hilbert Space $\mathcal{H}$ \cite{Baumgratz14}. Denote $\mathcal{I}$ the set of incoherent states. An incoherent state $\delta\in \mathcal{I}$ is of the form
\begin{equation}\label{IncoherentDefinition}
\delta=\sum_{i=1}^d\delta_i|i\rangle\langle i|.
\end{equation}
It is easily seen that $\mathcal{I}$ is convex and compact. From the Hahn-Banach theorem \cite{Edwards65}, there exists at least one coherence witness to detect a coherent state.

In analogy with entanglement witnesses \cite{Lewenstein00}, we give the following definitions about coherence. Given a coherence witness $W$, denote $D_{W}=\{\rho\geq0|\textrm{tr}\left(\rho W\right)<0\}$ the set of coherent states `` witnessed " by $W$.
Given two coherence witnesses $W_1$ and $W_2$, we say that $W_2$ is finer than $W_1$ if $D_{W_1}\subseteq D_{W_2}$, namely, if all the states witnessed by $W_1$ are also witnessed by $W_2$. We say that $W$ is an optimal coherence witness if there exist no other coherence witnesses which are finer than $W$.

Similar to the criteria for entanglement witnesses \cite{Lewenstein00}, for coherence we have the following conclusions on the conditions if a coherence witness is finer than another one, and a coherent witness is optimal.

{\textbf{Lemma 1:}} Assume that the witness $W_2$ is finer than the witness $W_1$.
Denote
\begin{equation}
\label{delta0}
\xi\equiv \inf_{\rho_1\in D_{W_1}} \left|
\frac{tr(W_2\rho_1)}{tr(W_1\rho_1)} \right|.
\end{equation}
We have

(a) If $tr(W_1\rho)=0$, then $tr(W_2\rho)\leq 0$;

(b) If $tr(W_1\rho)<0$, then $tr(W_2\rho) \leq
tr(W_1\rho)$;

(c) If $tr(W_1\rho)>0$, then $\xi\, tr(W_1\rho)\geq
tr(W_2\rho)$;

(d) $\xi\geq 1$;

(e) $D_{W_1}=D_{W_2}$ if and only if $W_1=W_2$.

{\em [Proof]:} Since $W_2$ is finer than $W_1$, we can use the
fact that $tr(W_2\rho)<0$ for all $\rho\geq 0$ such that $tr(W_1\rho)<0$.

{\em (a)} Assume that $tr(W_2\rho) > 0$. Take
any $\rho_1\in D_{W_1}$ so that for all $x\ge 0$, $\tilde \rho(x)\equiv \frac{1}{1+x}(\rho_1+x\rho) \in D_{W_1}$. Nevertheless, for sufficiently large $x$, $tr(W_2\tilde \rho(x))$
is positive, which cannot be true otherwise $\tilde \rho(x)\notin D_{W_2}$.

{\em (b)} Define $\tilde \rho=\frac{1}{1+|tr(W_1\rho)|}(\rho+|tr(W_1\rho)|\mathbb{I})$.
We have that ${\rm tr} (W_1\tilde{\rho})=0$. Using (a) we have
that $0\geq tr(W_2\rho)+|tr(W_1\rho)|$.

{\em (c)} Take $\rho_1\in D_{W_1}$ and define (unnormalized) $\tilde \rho=
tr(W_1\rho) \rho_1 + |tr(W_1\rho_1)| \rho$ such that $tr(W_1\tilde
\rho)=0$. Using (a) we have $|tr(W_1\rho_1)|tr(W_2\rho)\le
|tr(W_2\rho_1)|tr(W_1\rho)$. Dividing both sides by
$|tr(W_1\rho_1)|>0$ and $tr(W_1\rho)>0$ we obtain
\begin{equation}
\frac{tr(W_2\rho)}{tr(W_1\rho)} \le  \left|
\frac{tr(W_2\rho_1)}{tr(W_1\rho_1)} \right|.
\end{equation}
Taking the infimum with respect to $\rho_1\in D_{W_1}$ on the
right hand side of above equation we obtain the desired result.

{\em (d)} By {\em (b)}, it immediately follows that $\xi\geq1$.

{\em (e)} Denote the Hilbert-Schmidt inner product on $M_n$:  $\langle A, B \rangle=tr(AB^{\dag})$ and the H-S norm $\|A\|_2=[tr(AA^{\dag})]^{\frac{1}{2}}$.
If $tr(AB^{\dag})=\langle A, B \rangle=0$, we say $A\perp B$. If $tr(AB^{\dag})=\langle A, B \rangle \neq 0$, we say $A \not \perp B$. We require that
$\|W_1\|_2=\|W_2\|_2=1$.

We prove the only if part (the if part is trivial): if $D_{W_1}=D_{W_2}$, then $W_1=W_2$.
Suppose $W_1 \neq W_2$, we need to prove that there is a quantum state $\rho$, such that $tr(W_1\rho)=0$ but $tr(W_2 \rho)<0$, namely,
$\rho$ can detected by $W_2$, but not by $W_1$. Since $\|W_1\|_2=\|W_2\|_2=1$ and
$W_1 \neq W_2$, we can always find a state $\rho$ such that $\rho \perp W_1$ and $\rho \not \perp W_2$, that is, $tr(W_1\rho)=0$ and $tr(W_2 \rho)\neq 0$.
Hence, we can get that $tr(W_1\rho)=0$ and $tr(W_2 \rho)<0$ by (a).
\hfill $\Box$

As an example concerning the above proof of property (e), let us consider a 3-dimensional case:
$W_1 =\frac{1}{4} \left(
      \begin{array}{ccc}
        1 & 1 & 2\\
        1 & 1 & 0\\
        2 & 0 & 2\\
      \end{array}
    \right)$ and
$W_2 =\frac{1}{5} \left(
          \begin{array}{ccc}
            1 & 3 & 1\\
            3 & 1 & 1\\
            1 & 1 & 1\\
          \end{array}
        \right)$.
One can find that the quantum state
$\rho =\frac{1}{7} \left(
  \begin{array}{ccc}
    4 & -2 & -1\\
    -2 & 2 & 0\\
    -1 & 0 & 1\\
  \end{array}
\right)$ satisfies that
$tr(W_1\rho)=0$ and $tr(W_2\rho)=-\frac{1}{5}<0$.

By using Lemma 1 we can derive the following conclusions, see proofs in Appendix.

\textbf{ Corollary 1:} $W_2$ is finer than $W_1$ if and only if there exists a positive operator $P$ and
$1>\epsilon\geq 0$ such that $W_1=(1-\epsilon)W_2+\epsilon P$.


\textbf{ Corollary 2:} $W$ is optimal if and only if for all (unnormalized) coherent states $P$ and
$\epsilon>0$, $W'=(1+\epsilon)W-\epsilon P$ is not a coherence witness (it does not fulfill the condition (i)).

Corollary 1 and 2 tell us that $W$ is optimal if and only if when we subtract any positive operator from it, the resulting operator is not positive on incoherent states. In fact, from  Corollary 1 if all the diagonal elements of $W$ are zero, there would be no coherence witnesses which are finer than $W$, i.e., $W$ is optimal. On the other hand, from Corollary 2 if $W'$ is a coherence witness,
then $W$ is not optimal according to Corollary 1. Therefore, we have the following theorem:

\textbf{Theorem 1:} A coherence witness $W$ is optimal if and only if its diagonal elements are all zero.

Theorem 1 tells us that an optimal coherence witness $W$ with the condition (iii) ($tr(W)=1$) does not exist, as one cannot stop subtracting the positive operator (the incoherent state) until all diagonal elements become zero. Note that, following the definition of incoherent states and the Hahn-Banach theorem, we can also restrict the condition (i) $tr(W\delta)\geq0$ to that all diagonal elements of a coherence witness equal zero, and relax condition (ii) to $tr(W\pi)\neq 0$ (see the statement [50] of Ref. \cite{Napoli16} and Ref. \cite{Ren2017}). Although there intuitively exist traceless optimal coherence witnesses \cite{Napoli16,Ren2017}, to our knowledge, this fact has not been rigorously proven so far.

Generally, the construction of entanglement witnesses for a given entangled state is very difficult. The determination of entanglement witnesses for all entangled states is a nondeterministic polynomial-time hard problem \cite{Gurvits04,Doherty04,Hou10a}.
However, different from the construction of entanglement witnesses \cite{Guhne09,Horodecki09}, the construction of optimal coherence witnesses is rather easier for a given coherent state $\rho$, and the coherence witnesses can be experimentally implemented directly.

{\textbf{Theorem 2:}} For an arbitrary coherent state $\rho$, we can construct an optimal coherence witness $W_{\rho}=-\rho+\Delta(\rho)$ to detect the coherence of $\rho$, where $\Delta(\rho)=\sum_{i=0}^{d-1}\langle i|\rho|i\rangle|i\rangle\langle i|$ is the dephasing operation in the reference basis $\{|i\rangle\}_{i=0}^{d-1}$.

The proof is given in the Appendix.

\textit{Efficient detection of coherence -} It was thought that the coherence witnesses with nonzero trace can be friendlier in experimental implementation than traceless coherence witnesses, and can `set the scene for a practical verification of coherence', see the statement [50] of Ref. \cite{Napoli16}. Next we show that the optimal coherence witnesses with zero diagonal elements, compared the coherence witnesses with trace one, are experimentally friendlier in the sense of the statement [50] of Ref. \cite{Napoli16}. By Eq. (\ref{IncoherentDefinition}) and Lemma 1, we can construct a group of optimal witnesses to witness the real part or the imaginary part of any non-diagonal elements of a density matrix $\rho$,
\begin{equation}\label{GroupWitnesses}
\left\{
\begin{aligned}
W_{l\neq m}^R&=\frac{1}{2}(|l\rangle\langle m|+|m\rangle\langle l|),&\quad\quad\quad(a)\\
W_{l\neq m}^I&=\frac{1}{2}i(|l\rangle\langle m|-|m\rangle\langle l|),&\quad\quad\quad(b)
\end{aligned}
\right.
\end{equation}
where $W_{l\neq m}^R$ ($W_{l\neq m}^I$) can detect all the real (imaginary) parts of all non-diagonal elements of $\rho$. Hence a unified optimal coherence witness can be obtained, similar to \cite{Ren2017},
\begin{equation}\label{UnifiedForm}
W^U=\sum_{lm}(p_{l\neq m}^RW_{l\neq m}^R+p_{l\neq m}^IW_{l\neq m}^I),
\end{equation}
where $p_{l\neq m}^R$ and $p_{l\neq m}^I$ are real coefficients.

Let $\mathcal{D}(\mathbb{C}^d)$ be the convex set of density operators acting on a $d$-dimensional Hilbert space. The $d^2-d$ witnesses (\ref{GroupWitnesses}) can detect all the coherent quantum states $\rho\in\mathcal{D}(\mathbb{C}^d)$. These optimal coherence witnesses correspond to specific traceless generators of the standard special unitary group $SU(d)$ \cite{Christian07,Bertlmann08}.
Together with the operator $W_d=\sqrt{\frac{2}{d(d+1)}}(\sum_{l=0}^{d-1}|l\rangle\langle l|-d|d\rangle\langle d|)$, the witnesses in (\ref{GroupWitnesses}) can be employed for state tomography \cite{THEW02}. Therefore, one obtains a universal way for detecting quantum coherence based on the simplified state tomography from the optimal coherence witnesses.

We illustrate the state tomography in quantum two-level systems by Stokes parameters \cite{James2001}. Consider a set of four intensity measurements: (1) transmit $50\%$ of the incident radiation with a filter regardless of its polarization; (2) transmit only horizontally polarized light with a polarizer; (3) transmit only light polarized at $45^\circ$ to the horizontal with a polarizer; and (4) transmit only right-circularly polarized light. We obtain the number of the photons counted by a detector, which is proportional to the classical intensity in these four measurements: $n_0=\frac{N}{2}(\langle H|\rho|H\rangle+\langle V|\rho|V\rangle)=\frac{N}{2}(\langle R|\rho|R\rangle+\langle L|\rho|L\rangle)$; $n_1=N(\langle H|\rho|H\rangle)=\frac{N}{2}(\langle R|\rho|R\rangle+\langle R|\rho|L\rangle+\langle L|\rho|R\rangle+\langle L|\rho|L\rangle)$; $n_2=N(\langle D|\rho|D\rangle)=\frac{N}{2}(\langle R|\rho|R\rangle+\langle L|\rho|L\rangle-i\langle L|\rho|R\rangle+i\langle R|\rho|L\rangle)$; $n_3=N(\langle R|\rho|R\rangle)$, where $|H\rangle$, $|V\rangle$, $|D\rangle=(|H\rangle-|V\rangle)/\sqrt2=exp(i\pi/4)(|R\rangle+i|L\rangle)/\sqrt2$, and $|R\rangle=(|H\rangle-i|V\rangle)/\sqrt2$ represent photons polarized in the linear horizontal, linear vertical, linear diagonal, and right-circular senses, respectively. $\rho$ denotes the density matrix in $\mathbb{C}^2\otimes\mathbb{C}^2$ for the polarization degrees of the light, and $N$ is a constant depending on the detector efficiency and light intensity.

For single qubit $\rho$, its density matrix is of the form,
\begin{equation}
\rho=\frac{1}{2}\sum_{i=0}^{3}\frac{\mathcal{S}_i}{\mathcal{S}_0}\sigma_i,
\end{equation}
where $\mathcal{S}_0\equiv 2n_0$, $\mathcal{S}_1\equiv 2(n_1-n_0)$, $\mathcal{S}_2\equiv 2(n_2-n_0)$ and $\mathcal{S}_3\equiv 2(n_3-n_0)$ are the Stokes parameters, $\sigma_0\equiv \mathbb{I}$ is the identity operator, $\sigma_1\equiv \sigma_x\equiv W_{01}^R$, $\sigma_2\equiv\sigma_y\equiv W_{01}^I$, and $\sigma_3\equiv\sigma_z$ are Pauli operators.
A quantum state is incoherent only if both $\mathcal{S}_1$ and $\mathcal{S}_2$ are zero.
One can learn the quantum state whether or not it is coherent by comparing the
photon numbers $n_0$ and $n_1$, and $n_0$ and $n_2$. Generally, a $d$-dimensional state $\rho_d$ can be written as a linear combination of the generators of $SU(d)$ group,
\begin{equation}
\rho_d=\frac{1}{d}\sum_{j=0}^{d^2-1}r_j\lambda_j,
\end{equation}
where $\lambda_{(l-1)^2+2(m-1)}=W_{l\neq m}^R$ and $\lambda_{(l-1)^2+2m-1}=W_{l\neq m}^I$ are the off-diagonal generators
of the $SU(d)$ group and the coefficient $r_0$ is one for normalization.
Since the observables are orthogonal, the similar measurement approach as the qubit case apply,
see \cite{THEW02} for detailed tomography method.

Recently, based on the fact that a coherence witness $W$ satisfies the conditions (i) and (ii), a few experiments have been put forward \cite{Nie19,Wang17,Zheng18,Ringbauer18,Ma19}. Wang et. al. introduced a witness-observable method \cite{Wang17}. The target quantum state can be expressed as $\rho=\frac{1}{2}(\mathbb{I}+r_\rho(\sin\theta_\rho\cos\varphi_\rho\sigma_x
+\sin\theta_\rho\sin\varphi_\rho\sigma_y+\cos\theta_W\sigma_z))$. The qubit witness observable $W=\frac{a}{2}(\mathbb{I}+r_W(\sin\theta_W\cos\varphi_W\sigma_x+\sin\theta_W\sin\varphi_W\sigma_y
+\cos\theta_W\sigma_z))$, where $0< a\leq2$, $r_W\leq \frac{2}{a}-1$ and $-1\leq r_W\cos\theta_W\leq1$. The traversal searching parameters are $\varphi_W$, $\theta_W$, and $r_W$ (or $a$), respectively.

We can improve the witness-observable method given in Ref. \cite{Wang17}. From Theorem 1, for any quantum qubit state, the optimal coherence witness must be of the form,
\begin{equation}
W=\left[\begin{array}{cc} 0 & a-bi \\
                        a+bi & 0 \end{array}\right],
\end{equation}
where $a,b$ are real numbers. Such $W$ can be normalized as \cite{Ren2017}, $W=\cos\theta_W\sigma_x+\sin\theta_W\sigma_y$, where $\theta\in [0,2\pi)$.
Therefore, the traversal searching process can be reduced to one parameter $\theta$ from three parameters for optimal coherence witness \cite{Wang17}. Although the extreme value of $tr(W\rho)$ has been obtained experimentally at the point where the coherence witness is traceless, the general proof that an optimal coherence witness is traceless is first given here. Table \ref{tab1} gives a comparison between the optimal coherence witness (OCW) and non optimal coherence witness (NOCW).
Our tomographic witnessing approach and the state-tomography approach only need to count photons. While the interference-fringe approach and the witness-observable approach introduced in \cite{Wang17} need an ancillary state or sweeping parameters, see Table \ref{tab1}.

\begin{widetext}
\begin{table}[htbp]
\centering
\caption{Comparison between OCW and NOCW}
\begin{tabular}{c|c|c|c|c}
    \hline
    \hline
 & Interference-fringe & Witness-observable & State-tomography & Tomographic witnessing\\
    \hline
  An ancillary state  & {yes} & no&no&no\\
    \hline
  Sweeping parameters & yes  & yes &no&no\\
    \hline
  Counting photons & no  & no &yes&yes\\
      \hline
\end{tabular}
\label{tab1}
\end{table}
 \end{widetext}

We remark that our approach is efficient compared with the previous results from the witness-observable based method \cite{Wang17}. As our optimal witnesses have vanishing diagonal elements, no diagonal observables need to be measured.
For $d$-dimensional case, it means that the number of observables is $d^2-d$ in general,
compared with $d^2-1$ required in state tomography.
However, for the optimal case the number of observables required is just one, since we only verify if the quantum state has non-vanishing coherence. In average, consider $N+1$ $(N=d^2-d)$ quantum states $\{\rho_i\}_{i=0}^N$, where $\rho_i$ denotes a quantum state with $i$ real or imaginary zero non-diagonal entries. Suppose that each quantum state is randomly selected with probability $\frac{1}{N+1}$. Then the expected number of measurements is $E_N=\frac{1}{N+1}\sum_{i=0}^NE(\rho_i)$, where $E(\rho_i)=\sum_{m=1}^{i+1}m\cdot\frac{C_i^{m-1}}{C_N^{m-1}}\cdot\frac{N-i}{N-(m-1)}$ for $1\leq i\leq N-1$. Here
$C_i^{m-1}=\frac{i !}{(m-1) !(i-m+1) !}$ and
$C_N^{m-1}=\frac{N !}{(m-1) !(N-m+1) !}$
are combinatorial numbers. Then we can get that $E(\rho_0)=1$ for the best case $i=0$, and $E(\rho_N)=N$ for the worst case $i=N$. That is
\begin{equation}
E_N=\frac{1}{N+1}\{1+N+\sum_{i=1}^{N-1}\sum_{m=1}^{i+1}m\cdot\frac{C_i^{m-1}}{C_N^{m-1}}\cdot\frac{N-i}{N-(m-1)}\}.
\end{equation}
As an example, let us consider a 3-qubit Dicke state \cite{Dicke1954},
$|\psi\rangle=\sum_{i=0}^7|i\rangle$,
In this case, all the real (imaginary) parts of the non-diagonal entries of the density matrix are (nonzero) zero. The expected number of measurements in average is $E(\rho_{56})\approx 1.982$, much less than $d^2-d=56$.


\textit{The holographic measure -}
From the above analysis on witnessing both the real and imaginary parts of the non-diagonal entries
of a density matrix, we now present a new well defined measure of quantum coherence.

As the coherence are contributed by the non-diagonal entries of a density matrix $\rho$ under a fixed basis, an intuitive quantification of coherence would certainly be related to both the real and the imaginary parts of the off-diagonal entries of a quantum state. We have the following conclusion, see proof in Appendix.

{\textbf{Theorem 3:}} The following $\mathscr{C}_h(\rho)$ is a bona fide measure of coherence for a general quantum state $\rho$,
\begin{equation}\label{H_measure}
\mathscr{C}_h(\rho)=\sum_{l\neq m;l,m=1}^n|\rho_{lm}^R|+\sum_{l\neq m;l,m=1}^n|\rho_{lm}^I|,
\end{equation}
where $\rho_{lm}^R$ and $\rho_{lm}^I$ denote the real and the imaginary parts of $\rho_{lm}$, respectively.

We call $\mathscr{C}_h(\rho)$ a holographic measure of coherence as it involves both real and imaginary parts of all non-diagonal elements of the density matrix.
Clearly, $\mathscr{C}_h(\rho)=0$ if and only if $\rho$ is an incoherent state. The holographic measure can be expressed according to coherence witnesses.
Assigning $\frac{2\rho_{lm}^R}{|\rho_{lm}^R|}$ and $\frac{2\rho_{lm}^I}{|\rho_{lm}^I|}$ to $p_{lm}^R$ and $p_{lm}^I$ in (\ref{UnifiedForm}) according to the sign of the real and the imaginary parts of the entries of the density matrix, we have $\mathscr{C}_h(\rho)=tr(W^U\rho)$. Therefore, the measure is observable as it is just the expectation value of a coherence witness operator for any quantum state.

Besides the necessary conditions (C1), (C2a), (C2b) and (C3) \cite{Baumgratz14}, the measure of coherence $\mathscr{C}_h(\rho)$ also satisfies the following condition,

\begin{flalign}\label{C2c}
\text{(C2c)}\quad\mathscr{C}_h({\rho})\geq \mathscr{C}_h(\sum_i p_i|i\rangle\langle i|\otimes{\rho_i}),
~~~~~~~~~~~~~~~~~~
\end{flalign}
where $\rho=\sum_i p_i\rho_i$.

{\em [Proof]:} The holographic measure straightforwardly fulfil this additional constraint
as it satisfies (C2b), (C3) and $\mathscr{C}_h(|i\rangle\langle i|\otimes{\rho})\leq \mathscr{C}_h({\rho})$.
Under the $h$-norm, for any $|i\rangle\langle i|\in\mathcal{I}$ and matrix $\hat{M}$, one has
\begin{equation}\nonumber
\begin{split}
\bigl\||i\rangle\langle i|\otimes\hat{M}\bigr\|_{h}
&
=\sum_{j,k,l,m}
\Bigl|\bigl(|i\rangle\langle i|\otimes\hat{M}\bigr)_{(j,k),(l,m)}^R\Bigr|\\
&+\sum_{j,k,l,m}
\Bigl|\bigl(|i\rangle\langle i|\otimes\hat{M}\bigr)_{(j,k),(l,m)}^I\Bigr| \\
&=\sum_{j,k,l,m}
\delta_{j,i}\delta_{l,i}
|\hat{M}_{k,m}^R|+\sum_{j,k,l,m}
\delta_{j,i}\delta_{l,i}
|\hat{M}_{k,m}^I|\\
&=\sum_{k,m}
|\hat{M}_{k,m}^R|+\sum_{k,m}
|\hat{M}_{k,m}^I|,
\end{split}
\end{equation}
i.e., $\bigl\||i\rangle\langle i|\otimes({\rho}-{\rho}_{\text{diag}})\bigr\|_{h}= \bigl\|{\rho}-{\rho}_{\text{diag}}\bigr\|_{h}$. Therefore,
\begin{equation}\nonumber
\begin{split}
&\mathscr{C}_{h}\Bigl(\sum_ip_i|i\rangle\langle i|\otimes{\rho}_i\Bigr)\\
&\hspace{1cm}\stackrel{\text{(C3)}}{\le}
\sum_ip_i\mathscr{C}_{h}\bigl(|i\rangle\langle i|\otimes{\rho}_i\bigr)\\
&\hspace{1cm}\le \sum_ip_i\bigl\||i\rangle\langle i|\otimes{\rho}_i-|i\rangle\langle i|\otimes{\rho}_i^{\text{diag}}\bigr\|_{h}\\
&\hspace{1cm}=\sum_ip_i\bigl\|{\rho}_i-{\rho}_i^{\text{diag}}\bigr\|_{h}
=\sum_ip_i\mathscr{C}_{h}({\rho}_i)\\
&\hspace{1cm}\stackrel{\text{(C2b)}}{\le}\mathscr{C}_{h}({\rho}),
\end{split}
\end{equation}
which completes the proof.
\hfill $\Box$

The well-known $l_1$ norm measure $\mathscr{C}_{l_1}$ is defined by $\mathscr{C}_{l_1}(\rho)=\sum_{j\neq k}|\langle j|\rho|k\rangle|$. It is easy to see that $\mathscr{C}_h(\rho)\geq \mathscr{C}_{l_1}(\rho)\geq \frac{\sqrt2}{2}\mathscr{C}_h(\rho)$. For real density matrices, the holographic measure reduces to the $l_1$ norm measure.

The holographic measure has also an interesting relationship with the robust of coherence.
The robust of coherence $\mathscr{C}_\mathcal{R}(\rho)$ \cite{Piani16,Napoli16} of a quantum state $\rho \in {\mathcal{D}}(\mathbb{C}^d)$ is defined as
\begin{equation}\label{ROC}
 \mathscr{C}_\mathcal{R}(\rho)= \min_{\tau \in {\mathscr{D}}(\mathbb{C}^d)} \left\{ s\geq 0\ \Big\vert\ \frac{\rho + s\ \tau}{1+s} =\delta \in {\mathcal{I}}\right\}.
\end{equation}
It is found that the expectation value of any witness $W$ provides a quantitative lower bound of $\mathscr{C}_\mathcal{R}$,
\begin{eqnarray}\label{maxValue}
\mathscr{C}_\mathcal{R}(\rho)\geq \text{max}\{0,-tr(W\rho)\}
\end{eqnarray}
$\forall\, W$ such that $\triangle(W)\geq0$ and $W\leq \mathbb{I}$.
Interestingly, given a state $\rho$, there always exists an optimal witness $W^*$ characterized by $\triangle(W^*)=0$, which saturates the inequality $(\ref{maxValue})$.

\textbf{Theorem 4:} For any state $\rho\in \mathscr{D}(\mathbb{C}^d)$ it holds that
\begin{equation}\label{thm4}
\mathscr{C}_h(\rho)=\mathscr{C}_\mathcal{R}(\rho)\mathscr{C}_h(\tau)
~~\text{and}~~\mathscr{C}_h(\tau)\leq1,
\end{equation}
where $\tau$ is the state minimizing $s$ in (\ref{ROC}).

 {\em [Proof]:} From (\ref{ROC}) we have $\rho+\mathscr{C}_\mathcal{R}(\rho)\tau=(1+\mathscr{C}_\mathcal{R}(\rho))\delta$. Since $\mathscr{C}_h(\delta)=0$ we get $\mathscr{C}_h(\rho+\mathscr{C}_\mathcal{R}(\rho)\tau)=0$.
As $\mathscr{C}_h(\rho)=\mathscr{C}_h(c\mathbb{I}-\rho)$ for any quantum state $\rho$ and $c>0$ such that $c\mathbb{I}-\rho$ is positive, we obtain $\mathscr{C}_h(\rho)=\mathscr{C}_\mathcal{R}(\rho)\mathscr{C}_h(\tau)$. \hfill $\Box$

Theorem 4 gives a direct relation between the holographic measure of coherence and the robustness of coherence. Since the robustness of coherence quantifies the advantage enabled by a quantum state in a phase discrimination task, holographic measure of coherence plays also essential roles in characterizing
phase discriminations.

\textit{Conclusions and discussions -} We have shown that the optimal coherence witness
with trace does not exist. Based on this fact, we have presented a more efficient tomographic witnessing method for detecting coherence in experiments.
From the analysis on optimal coherence witnesses, we have put forward a bona fide
measure based on the sum of absolute values of the real and imaginary parts of non-diagonal
elements of a density matrix, and analyzed the relations among the $l_1$ norm coherence, robustness of coherence and holographic measure of coherence. Our results may highlight further investigations on the theory of quantum coherence and its applications.
\smallskip

\begin{acknowledgments}
\textit{Acknowledgments} We are grateful to Zi-Wen Liu, Xiaoqi Zhou, Xiao Yuan, Tristan Farrow, Jinzhao Sun, Zong Wang and Gang-Gang Cao for helpful discussions, especially Paul Kairys for pointing out the mistakes in the proof of Lemma 1 and Theorem 3 in the original manuscript. Wang thanks Vlatko Vedral for his kind hospitality at University of Oxford where part of this work was initiated when Wang was there as a visiting scholar.
This work is supported by the National Natural Science Foundation of China under Grant Nos. 62072119, 61672007, 11675113 and 12075159, Guangdong Basic and Applied Basic Research Foundation under Grant No. 2020A1515011180, Shenzhen Institute for Quantum Science and Engineering, Southern University of Science and Technology (Grant Nos. SIQSE202005, SIQSE202001), Natural Science Foundation of Shanghai (Grant No. 20ZR1426400),
the Key Project of Beijing Municipal Commission of Education (Grant No. KZ201810028042), Beijing Natural Science Foundation (Z190005), the Academician Innovation Platform of Hainan Province, and Academy for Multidisciplinary Studies, Capital Normal University.
\end{acknowledgments}

B.-H. Wang and S.-Q. Zhou contributed equally to this work.






\newpage

\section{APPENDIX}

\subsection{Proof of Corollary 1}

(If) For all $\rho\in D_{W_1}$ we have that
$0>tr(W_1\rho)= (1-\epsilon)tr(W_2\rho)+\epsilon tr(P\rho)$,
which implies $tr(W_2\rho)<0$ and hence $\rho\in D_{W_2}$.

(Only if) We have
$D_{W_1}\subseteq D_{W_2}$ if $W_2$ is finer than $W_1$. If $D_{W_1}=D_{W_2}$, then Lemma 1 (e) gives rise to that $W_1=W_2$ (i.e., $\epsilon=0$).
If $D_{W_1}\subset D_{W_2}$, we have $W_1 \neq W_2$ due to $D_{W_1}\neq D_{W_2}$ by using Lemma 1 (e). Hence $tr(W_2\rho)\neq tr(W_1\rho)$. For all $\rho \in D_{W_1}$, we have $tr(W_1\rho)<0$. Combining with Lemma 1 (b), we get that $tr(W_2\rho)<tr(W_1\rho)$ and $\xi > 1$. Denote $P=(\xi-1)^{-1}(\xi W_1-W_2)$ and $\epsilon=1-1/\xi>0$.
We have that $W_1=(1-\epsilon)W_2+\epsilon P$. It only remains
to be shown that $P\ge 0$. But this follows from Lemma 1 (a-c) and the
definition of $\xi$. One easily verifies that $P$ is either not finer than $W_1$ or it is incoherent.
\hfill $\Box$

\subsection{Proof of Corollary 2}

(If) According to Corollary 1, there is no coherence witness which is
finer than $W$. Therefore, $W$ is optimal.

(Only if) If $W'$ is a coherence witness, then according to Corollary 1 $W$ is not optimal.
\hfill $\Box$

\subsection{Proof of Theorem 2}

Clearly, $W_{\rho}$ is Hermitian and not positive. The diagonal elements of $W_{\rho}$ are 0, and $tr(W_{\rho}\delta)\geq0$ (factually $tr(W_{\rho}\delta)=0$) for any incoherent state $\delta$.

Since $\sum_{i=0}^{d-1}\langle i|\rho|i\rangle|i\rangle\langle i|$ is a matrix with the same diagonal elements as $\rho$ and other elements 0, we have $tr(\rho^2)>tr(\rho\sum_{i=0}^{d-1}\langle i|\rho|i\rangle|i\rangle\langle i|)$ and
$$
tr(\rho W_{\rho})=-tr(\rho^2)+tr(\rho\sum_{i=0}^{d-1}\langle i|\rho|i\rangle|i\rangle\langle i|)<0.
$$
\hfill $\Box$

\subsection{Proof of Theorem 3}

A proper coherence measure $\mathscr{C}(\rho)$ for a quantum state $\rho$ should satisfy the following conditions:

(C1) Non-negativity. $\mathscr{C}(\rho)\geq0$ for any quantum state $\rho$, and $\mathscr{C}(\rho)=0$ if and only if $\rho$ is an incoherent state.

(C2a) Monotonicity. $\mathscr{C}(\rho)\geq \mathscr{C}(\Phi(\rho))$ for all incoherent completely positive and trace-preserving (ICPTP) maps $\Phi(\rho)=\sum_n\hat{K}_n\rho \hat{K}_n^\dag$, where $\{\hat{K}_n\}$ is a set of Kraus operators, $\sum_n\hat{K}_n^\dag \hat{K}_n=\mathbb{I}$ and $\hat{K}_n\mathcal{I}\hat{K}_n^\dag\subset \mathcal{I}$.

(C2b) Strong monotonicity (under selective measurements on average). $\mathscr{C}(\rho)\geq\sum_np_n\mathscr{C}(\rho_n)$, where $\rho_n=\frac{\hat{K}_n\rho \hat{K}_n^\dag}{p_n}$, $p_n=tr(\hat{K}_n\rho \hat{K}_n^\dag)$, $\sum_n\hat{K}_n^\dag \hat{K}_n=\mathbb{I}$ and $\hat{K}_n\mathcal{I}\hat{K}_n^\dag\subset \mathcal{I}$.

(C3) Convexity. $\sum_np_n \mathscr{C}(\rho_n)\geq \mathscr{C}(\sum_n p_n\rho_n)$ for any ensemble of $\{p_n,\rho_n\}$ of a state $\rho$.

For the convinience of proof, we define a so-called holographic matrix norm.
Denote $M_n$ the $n\times n$ complex matrices $M_{n,n}(C)$.
We define the holographic norm of $A\in M_n$ by
\begin{equation}\label{norm}
\parallel A\parallel_h=\sum_{l,m=1}^n|a_{lm}^R|+\sum_{l,m=1}^n|a_{lm}^I|,
\end{equation}
where $a_{lm}^R$ and $a_{lm}^I$ are the real and the imaginary parts of $a_{lm}$, respectively.
We clarify that (\ref{norm}) is not a true matrix norm but a pseudo matrix norm.
Recall that a matrix norm $\|\cdot\|$ is a function satisfying the following properties: for $A,B\in M_n$,
 \begin{eqnarray}\nonumber
 &\text { (1) }& \|A\| \geq 0,~ \|A\|=0 \text { iff } A=0 \text { (Nonnegative) } \\\nonumber
& \text { (2) }& \|c A\|=|c|\|A\| \text {for all complex c} \text {(Homoheneous)}\\\nonumber
 &\text { (3) }& \|A+B\| \leq\|A\|+\|B\| \text { (Triangle inequality) } \\\nonumber
 &\text { (4) }& \|A B\| \leq\|A\|\|B\| \text { (Submultiplicative). }
 \end{eqnarray}
One can prove that $\|A\|_{h}$ defined in (\ref{norm}) satisfies the properties (1), (3) and (4), as well as the modified property (2a): $\|c A\|_{h}=|c|\|A\|_{h}$ for all real and pure imaginary number $c$, but not for arbitrary complex scalars $c$. We remark that we do not use this property for arbitrary complex scalars $c$ in our proof. We call $\|A\|_{h}$ a pseudo matrix norm.

Properties (1), (2a) and (3) are easy to prove.
We prove the property (4) as follows:
\begin{align}
&\|AB\|_{h} \notag \\
\leq&\sum_{l,m=1}^{n}|(\sum_{j=1}^{n} a_{lj}b_{jm})^{R}|+\sum_{l,m=1}^{n}|(\sum_{j=1}^{n}a_{lj}b_{jm})^{I}|\notag \\
=&\sum_{l,m=1}^{n}|\sum_{j=1}^{n}(a_{lj}^{R} b_{jm}^{R}-a_{lj}^{I}b_{jm}^{I})|+\sum_{l,m=1}^{n}|\sum_{j=1}^{n}(a_{lj}^{R}b_{jm}^{I}+a_{lj}^{I}b_{jm}^{R})|\notag \\
\leq&\sum_{l,m=1}^{n}\sum_{j=1}^{n}(|a_{lj}^{R}| |b_{jm}^{R}|+|a_{lj}^{I}| |b_{jm}^{I}|+|a_{lj}^{R}||b_{jm}^{I}|+|a_{lj}^{I}||b_{jm}^{R}|)\notag \\
=&\sum_{l,m=1}^{n}\sum_{j=1}^{n} (|a_{lj}^{R}|+ |a_{lj}^{I}|)(|b_{jm}^{R}|+|b_{jm}^{I}|)\notag \\
\leq&(\sum_{l,j=1}^{n}(|a_{lj}^{R}|+|a_{lj}^{I}|))(\sum_{j,m=1}^{n}(|b_{jm}^{R}|+|b_{jm}^{I}|))\notag \\\notag
=&\|A\|_{h}\|B\|_{h}.
\end{align}

Now we prove that the holographic measure of coherence $\mathscr{C}_h(\rho)$
satisfies the required conditions.

(C1) Obviously,  $\mathscr{C}_{h}(\rho)\geq0$ for any quantum state $\rho$ and $\mathscr{C}_{h}(\rho)=0$ if and only if $\rho$ is incoherent.

(C2\lowercase{b}) A general state can be written as
$\rho=\rho^R+i\rho^I$, where $\rho^R$ and $\rho^I$ are the real and imaginary parts of $\rho$, respectively. The holographic norm of coherence of $\rho$ can be written as
$$
\mathscr{C}_{h}(\rho)=\sum_{i, j  \atop i\neq j}|[\rho^R]_{i,j}|+\sum_{i, j  \atop i\neq j}|[\rho^I]_{i,j}|.
$$
Therefore, we have
\begin{align}\label{21}
  \sum_np_n\mathscr{C}_{h}({\rho}_n)=&\sum_{n}p_{n}\sum_{i, j  \atop i\neq j}(|[\rho_{n}^{R}]_{i,j}|+|[\rho_{n}^{I}]_{i,j}|)\notag \\
=&\sum_{n}p_{n}\sum_{i, j  \atop i\neq j}|[\rho_{n}^{R}]_{i,j}|+\sum_{n}p_{n}\sum_{i, j  \atop i\neq j}|[\rho_{n}^{I}]_{i,j}|\notag \\
=&\sum_{n}\sum_{i, j  \atop i\neq j}|p_{n}[\rho_{n}^{R}]_{i,j}|+\sum_{n}\sum_{i, j  \atop i\neq j}|p_{n}[\rho_{n}^{I}]_{i,j}|\notag \\
=&\sum_{n}\sum_{i, j  \atop i\neq j}|[\hat{K}_{n}\rho_{n}^{R}\hat{K}_{n}^{\dag}]_{i,j}|+\sum_{n}\sum_{i, j  \atop i\neq j}|[\hat{K}_{n}\rho_{n}^{I}\hat{K}_{n}^{\dag}]_{i,j}|\notag \\
=&\sum_{n}\sum_{i, j  \atop i\neq j}|\sum_{k,l}[\hat{K}_{n}]_{i,k}[\rho_{n}^{R}]_{k,l}[\hat{K}_{n}^{\dag}]_{l,j}|\notag \\
&+\sum_{n}\sum_{i, j  \atop i\neq j}|\sum_{k,l}[\hat{K}_{n}]_{i,k}[\rho_{n}^{I}]_{k,l}[\hat{K}_{n}^{\dag}]_{l,j}|.
\end{align}

Note that for $k=l$ in (\ref{21}), we have
$\sum_{k}[\hat{K}_{n}]_{i,k}\rho_{k,k}^{R}[\hat{K}_{n}^{\dag}]_{k,j}=\sum_{k}\langle i|\hat{K}_{n}|k\rangle\rho_{k,k}^{R}\langle k|\hat{K}_{n}^{\dag}|j\rangle=\langle i|\hat{K}_{n}\Delta(\rho^R)\hat{K}_{n}^{\dag}|j\rangle$,
where $\Delta(\rho^R)=\sum_{k}\rho_{k,k}^{R}|k\rangle\langle k|$. Due to the fact that $\Delta(\rho^{R})\in \mathcal{I}$ and $\hat{K}_{n} \Delta(\rho^{R}) \hat{K}_{n}^{\dagger}\in \mathcal{I}$, we obtain $\sum_{k}[\hat{K}_{n}]_{i,k}\rho_{k,k}^{R}[\hat{K}_{n}^{\dag}]_{k,j}=\delta_{i,j}(\hat{K}_{n}\Delta(\rho^R)\hat{K}_{n}^{\dag})$.
Hence, for the case $i\ne j$, we only need to consider the case of $k\ne l$.
Then the term of the real part in (\ref{21}) can be written as
\begin{align}\label{22}
&\sum_{n}\sum_{i, j  \atop i\neq j}|\sum_{k, l}[\hat{K}_{n}]_{i, k}[\rho_{n}^{R}]_{k, l}[\hat{K}_{n}^{\dag}]_{l, j}|\notag \\
=&\sum_{n}\sum_{i, j  \atop i\neq j}|\sum_{k, l \atop k\neq l}[\hat{K}_{n}]_{i,k}[\rho_{n}^{R}]_{k,l}[\hat{K}_{n}^{\dag}]_{l,j}|\notag \\
\le&\sum_{k, l \atop k\neq l}|[\rho_{n}^{R}]_{k,l}|\sum_{n}\sum_{i, j  \atop i\neq j}|[\hat{K}_{n}]_{i,k}[\hat{K}_{n}^{\dag}]_{l,j}|\notag \\
\le&\sum_{k, l \atop k\neq l}|[\rho_{n}^{R}]_{k,l}|\sum_{n}\sum_{i}|[\hat{K}_{n}]_{i,k}|\sum_{j}|[\hat{K}_{n}^{\dag}]_{l,j}|.
\end{align}
It is easy to see that
\begin{align}
&|\sum_{n}\sum_{i}|[\hat{K}_{n}]_{i,k}|\sum_{j}|[\hat{K}_{n}^{\dag}]_{l,j}|\notag \\
\le&\sqrt{\sum_{n}(\sum_{i}|[\hat{K}_{n}]_{i,k}|)^{2}\sum_{m}(\sum_{j}|[\hat{K}_{n}^{\dag}]_{l,j}|)^{2}}.
\end{align}
Sincce
\begin{align}
&\sum_{n}(\sum_{i}|[\hat{K}_{n}]_{i,k}|)^{2}\notag \\
=&\sum_{n}\sum_{i,j}|[\hat{K}_{n}]_{i,k}[\hat{K}_{n}^{\dag}]_{k,j}|\notag \\
=&\sum_{n}\sum_{i,j}|\langle i|\hat{K}_{n}|k\rangle\langle k|\hat{K}_{n}^{\dag}|j\rangle|\notag \\
=&\sum_{n}\sum_{i}|\langle i|\hat{K}_{n}|k\rangle\langle k|\hat{K}_{n}^{\dag}|j\rangle|\notag \\
=&\sum_{n}\sum_{i}|\langle k|\hat{K}_{n}^{\dag}|j\rangle\langle i|\hat{K}_{n}|k\rangle|\notag \\
=&1,
\end{align}
and similarly, $\sum_{m}(\sum_{j}|[\hat{K}_{n}^{\dag}]_{l,j}|)^{2}=1$, we obtain
\begin{equation}\label{25}
|\sum_{n}\sum_{i}|[\hat{K}_{n}]_{i,k}|\sum_{j}|[\hat{K}_{n}^{\dag}]_{l,j}|\le1.
\end{equation}

From (\ref{22}) and (\ref{25}), we have
\begin{equation}\label{26}
\sum_{n}\sum_{i, j  \atop i\neq j}|\sum_{k,l}[\hat{K}_{n}]_{i,k}[\rho_{n}^{R}]_{k,l}[\hat{K}_{n}^{\dag}]_{l,j}|\le\sum_{k, l \atop k\neq l}|[\rho_{n}^{R}]_{k,l}|.
\end{equation}
In a similar way, one can aslo prove that
\begin{equation}\label{27}
\sum_{n}\sum_{i, j  \atop i\neq j}|\sum_{k,l}[\hat{K}_{n}]_{i,k}[\rho_{n}^{I}]_{k,l}[\hat{K}_{n}^{\dag}]_{l,j}|\le\sum_{k, l \atop k\neq l}|[\rho_{n}^{I}]_{k,l}|.
\end{equation}
Substituting (\ref{26}) and (\ref{27}) into (\ref{21}), we get
\begin{equation}
\sum_{n}p_{n}\mathscr{C}_{h}(\rho_{n})\le\sum_{k, l \atop k\neq l}|[\rho^R]_{i,j}|+\sum_{k, l \atop k\neq l}|[\rho^I]_{i,j}|=\mathscr{C}_{h}(\rho),
\end{equation}
which completes the proof.

(C3) That the holographic norm of coherence may only decrease under mixing can be proved
directly by the triangle inequality $\parallel A+B\parallel_h\leq\parallel A\parallel_h+\parallel B\parallel_h$ satisfied by the holographic norm.

(C2a) By (C2b) and (C3), we have
$$
\mathscr{C}_h(\Phi({\rho}))=\mathscr{C}_h(\sum_np_n{\rho}_n)\stackrel{\text{(C3)}}{\le}\sum_np_n\mathscr{C}_h({\rho}_n)
\stackrel{\text{(C2b)}}{\le}\mathscr{C}_{h}({\rho}),
$$
where ${\rho}_n=\hat{K}_n{\rho} \hat{K}_n^\dag/p_n$ and $p_n=tr(\hat{K}_n{\rho} \hat{K}_n^\dag)$.
\hfill $\Box$

\end{document}